\documentstyle[11pt,epsf]{article}
\font\tenbf=cmbx9

\font\twelbf=cmbx12
\font\tenrm=cmr9

\font\tenit=cmti9
\font\sc=cmr12

\font\lrm=cmr9

\newcommand{\lle}{\mbox{$\langle$}}
\newcommand{\rle}{\mbox{$\rangle$}}

\newcommand{\bfsi}{\mbox{\boldmath$\sigma$}}
\newcommand{\bfep}{\mbox{\boldmath$\varepsilon$}}

\newcommand{\bfcK}{\mbox{\boldmath$\cal K$}}
\newcommand{\bfcL}{\mbox{\boldmath$\cal L$}}

\newcommand{\bfg}{\mbox{\boldmath$\bf g$}}

\newcommand{\bft}{\mbox{\boldmath$\bf t$}}
\newcommand{\bfs}{\mbox{\boldmath$\bf s$}}
\newcommand{\bfu}{\mbox{\boldmath$\bf u$}}

\newcommand{\bfx}{\mbox{\boldmath$\bf x$}}
\newcommand{\bfy}{\mbox{\boldmath$\bf y$}}
\newcommand{\bfz}{\mbox{\boldmath$\bf z$}}
\newcommand{\bfA}{\mbox{\boldmath$\bf A$}}

\newcommand{\bfE}{\mbox{\boldmath$\bf E$}}

\newcommand{\bfI}{\mbox{\boldmath$\bf I$}}

\newcommand{\bfL}{\mbox{\boldmath$\bf L$}}
\newcommand{\bfN}{\mbox{\boldmath$\bf N$}}
\newcommand{\bfP}{\mbox{\boldmath$\bf P$}}

\newcommand{\bfY}{\mbox{\boldmath$\bf Y$}}

\newcommand{\bfU}{\mbox{\boldmath$\bf U$}}

\newcommand{\bfR}{\mbox{\boldmath$\bf R$}}
\newcommand{\bfK}{\mbox{\boldmath$\bf K$}}
\newcommand{\bfM}{\mbox{\boldmath$\bf M$}}
\newcommand{\bfT}{\mbox{\boldmath$\bf T$}}
\newcommand{\bfS}{\mbox{\boldmath$\bf S$}}

\newcommand{\bfdel}{\mbox{\boldmath$\delta$}}

\newcommand{\bfcI}{\mbox{\boldmath$\cal I$}}

\newcommand{\bfcR}{\mbox{\boldmath$\cal R$}}

\newcommand{\bftau}{\mbox{\boldmath$\tau$}}

\newcommand{\BB}{\begin{equation}}
\newcommand{\EE}{\end{equation}}
 
\newcommand{\BBEQ}{\begin{eqnarray}}
\newcommand{\EEEQ}{\end{eqnarray}}

\begin{document}

\vspace{20pt}
\centerline {\twelbf On some background of micromechanics}
\vspace{10pt} 

\centerline {\twelbf of random structure matrix composites}

\vspace{40pt}

\centerline{\sc Valeriy  A. Buryachenko
\footnote{\tenrm  Corresponding author: Tel.: +39 070 675-5401; fax 
+39 070 6755418; email:Buryach@aol.com}}

\vspace{15pt}
\centerline{\tenit Department of Structural Engineering, 
University of Cagliari,
09124 Cagliari,  Italy}

\
\vspace{35pt}

\noindent{\vrule height 0.003 in width 16.cm}

 \noindent{\tenbf Abstract:}
\vspace{15pt}

 \noindent{\baselineskip=8pt
 {\tenrm
We consider a linearly elastic composite medium, which consists of a homogeneous 
matrix containing statistically inhomogeneous random set of heterogeneities
and loaded by inhomogeneous remote loading. 
The new general integral equation is obtained by a centering procedure without any auxiliary assumptions such as, e.g., effective field hypothesis implicitly exploited in the known centering methods. 
The method makes it possible to abandon the basic concepts of classical micromechanics such as effective field hypothesis, and the hypothesis of ``ellipsoidal symmetry". 
The results of this abandonment leads to detection of some fundamentally new effects that is impossible in the framework of a classical background of micromechanics. \par}}
\vspace{10pt}

\noindent
{\lrm Keywords: A. microstructures, B. inhomogeneous material, B.
 elastic material.}
 
\smallskip
\centerline{\vrule height 0.003 in width 16.cm}

\vspace{15pt}

\noindent {\bf 1 Introduction}
\bigskip

The final goals of micromechanical research of composites
involved in  a prediction of both the overall effective properties 
 and statistical moments of stress-strain fields are based on the 
approximate solution of exact initial  general integral equations
connecting the random stress fields at the point being 
considered and the surrounding points. These equations are well-known for statistically homogeneous composite materials subjected to homogeneous boundary conditions. In the current paper, these known equations are generalized to the case of inhomogeneity of both the statistical microstructure and applied loading. 
The method is based on a centering procedure of subtraction from both sides of a known initial integral equation the statistical averages obtained without any auxiliary assumptions such as, e.g., effective field hypothesis implicitly exploited in the known centering methods.

A considerable
number of methods are known in the linear elasticity
theory of composites.  
Appropriate, but by no means exhaustive,
references are provided by the book by Buryachenko [1]. 
 Nowadays, it appears that variants of
the effective medium by Kr\"oner [2] and mean
field methods by Mori and Tanaka [3] are the
most popular and widely used methods. 
Recently, a new method has become known, the multiparticle effective
field method (MEFM) (see [1]). The
MEFM is based on the theory of functions of random
variables and Green's functions. Within this method
one constructs a hierarchy of statistical moment equations
for conditional averages of the stresses in the
inclusions. The hierarchy is then cut by introducing
the notion of an effective field. This way the interaction
of different   inclusions is taken into account.
Buryachenko [1] has demonstrated that the
MEFM includes as particular cases the well-known
methods of mechanics of strongly heterogeneous media.

However, all mentioned methods are based on the effective field hypothesis (EFH, even if the term ``{\it effective field hypothesis}" was  not indicated) according to which each inclusion is located inside a homogeneous so–called effective field (see for references [1]). Effective field hypothesis is apparently the most fundamental, most prospective, and most exploited concept of micromechanics. This concept has directed a development of micromechanics over the last sixty years and made a contribution to their progress incompatible with any another concept.
The idea of effective field dating back to Mossotti [4]  was added by 
the hypothesis of ``{\it ellipsoidal symmetry}" for the distribution of inclusions attributed to Willis [5]. 
However, we will show in this paper that the EFH (also called the hypothesis {\bf H1}) is a central one and other concept plays a satellite role providing the conditions for application of the EFH. Moreover, we will show that all mentioned hypotheses are not really necessary and can be relaxed.

The outline of the study is as follow. 
In Section 2 we present the basic equation of thermoelasticity, 
notations, and statistical description of the composite microstructure.
The new general integral equations  are proposed in Section 3 for the case of statistically 
inhomogeneous structures of composite materials.  These equations are obtained by a centering procedure of subtraction from both sides of a known initial integral equation the statistical averages obtained without any auxiliary assumptions such as, e.g., EFH implicitly exploited in the known centering methods. The new general integral equation is compared with the known ones. In Section 4 we recall the basic concepts defining the background of classical micromechanics. Explicit formulae for both effective elastic moduli and strain concentrator factor are presented.
The new  general integral equations presented in Section 5 through the operator form of the particular solutions for one heterogeneous in the infinite matrix subjected to inhomogeneous effective field.  
This equation is solved by the iteration method in the framework of the quasi-crystallite approximation but without basic hypotheses of classical micromechanics such as both the EFH and ``{\it ellipsoidal symmetry}" assumption. 
In Section 5 we qualitatively explain the advantages of the new approach with respect to the classic ones and demonstrate the corrections of popular propositions obtained in the framework of the old background of micromechanics. 

\vspace{15pt}
\noindent{\bf 2 
Description of the mechanical properties and geometrical structure of the components}
\bigskip

\setcounter{equation}{0}
\renewcommand{\theequation}{2.\arabic{equation}}
Let a linear   elastic body occupy an open bounded domain  $w\subset R^d$
 with a smooth boundary $\Gamma$ and with a characteristic
function $W$ and space dimensionality $d$
 ($d=2$ and $d=3$ for 2-$D$ and 3-$D$ problems, respectively).
The domain $w$ contains  a homogeneous matrix $v^{(0)}$ and
a statistically inhomogeneous set $X=(v_i,V_i,\bfx_i)$ of inclusions $v_i$
with characteristic functions $V_i$ and centers $\bfx_i$.
It is assumed that the inclusions can be grouped into
    component (phase) $v^{(1)}$ with
 identical  mechanical and geometrical properties (such as the
 shape, size,
orientation, and microstructure of inclusions).
For the sake of definiteness, in the 2-$D$ case we will consider a
plane-strain problem.  At first no restrictions are imposed on the
  elastic symmetry of the phases or on the geometry of the inclusions
\footnote{It is known that for 2-$D$ problems the plane-strain state
is only possible for material symmetry no lower than orthotropic
(see e.g. [6])
that will be assumed hereafter in 2-$D$ case.}.

We will consider the local basic equations of  elastostatics of composites
\BBEQ
\nabla\bfsi(\bfx)&=&{\bf 0},   \\ 
\bfsi(\bfx)&=&\bfL(\bfx)\bfep(\bfx), \ \
{\rm or}\ \ \
\bfep(\bfx)=\bfM(\bfx)\bfsi(\bfx),  \\ 
\bfep(\bfx)&=&[\nabla \otimes {\bf u}+(\nabla \otimes
{\bf u})^{\top}]/2,\ \
{\rm Inc}\bfep(\bfx)={\bf 0},  
\EEEQ
where $\otimes$ denotes tensor product,
 and $(.)^{\top}$  denotes matrix transposition.
${\bf L(x)}$ and ${\bf M(x) \equiv L(x)}^{-1}$
are the known stiffness and  compliance
fourth-order tensors, and the common notation for contracted products
has  been employed.

In particular, for isotropic constituents the stiffness
tensor $\bfL$ is given in terms of the local bulk modulus $k$ and the 
shear modulus $\mu$: 
$\bfL=(dk,2\mu)\equiv dk\bfN_1+2\mu\bfN_2$,
${\bf N}_1=\bfdel\otimes\bfdel/d, \ {\bf N}_2={\bf I-N}_1$ $(d=2\ {\rm or}\ 3$);
$\bfdel$ and $\bfI$ are the unit second-order and fourth-order tensors.
 The tensors ${\bf g}$ $({\bf g=L, M})$
of material properties are piecewise constant and
 decomposed as ${\bf g\equiv g}^{(0)}+{\bf g}_1({\bf x})=\bfg^{(0)}+
\bfg_1^{(1)}(\bfx)$
where $\bfg^{(0)}=$const, ${\bf g}(\bfx)\equiv\bfg^{(0)}$ at $\bfx\in v^0$ and
$\bfg_1^{(1)}(\bfx)\equiv \bfg_1^{(1)}$ is a homogeneous function of the
$\bfx\in v^{(1)}$:
\BB
\bfL^{(1)}_1(\bfx)=\bfL^{(1)}_1\equiv{\rm const}.\ {\rm at} \ 
\bfx\in v^{(1)}. 
\EE
 The upper index of the material properties tensor put in parentheses shows the number of the respective constituent. 
The upper index $^{(m)}$ indicates the
components and the lower index $i$ indicates the individual
inclusions; $v^{(0)}=w\backslash v$, $ v\equiv v^{(1)},
\ V(\bfx)=V^{(1)}=\sum V_i(\bfx)$, and $V^{(1)}(\bfx)$ and $V_i(\bfx)$ are the
indicator functions of $v^{(1)}$ and $v_i$, respectively.

The boundary conditions at the interface boundary  will be considered
together with the mixed boundary conditions on $\Gamma$
with the unit outward normal ${\bf n}^{\Gamma}$
\BBEQ
{\bf u}(\bfx)&=&{\bf u}^{\Gamma}(\bfx),\ \ \ \bfx\in \Gamma_u,  \\
\bfsi(\bfx){\bf n}^{\Gamma}(\bfx)&=&{\bf t}^{\Gamma}(\bfx), \ \ \ \bfx\in \Gamma_t,
\EEEQ
where $\Gamma_u$ and $\Gamma_t$ are prescribed displacement and traction
boundaries  such that $\Gamma_u\cup\Gamma_t=
\Gamma,\ \Gamma_u\cap\Gamma_t=\emptyset$.
${\bf u}^{\Gamma}(\bfx)$ and ${\bf t}^{\Gamma}(\bfx)$ are, respectively,
 prescribed
displacement on $\Gamma_u$ and traction on $\Gamma_t$.
Of special practical interest are the homogeneous boundary conditions
\BBEQ
\bfu^{\Gamma}(\bfx)&=&\bfep^{\Gamma}\bfx,\ \bfep^{\Gamma}\equiv{\rm const.},\
\bfx\in\Gamma,
\\
{\bf t}^{\Gamma}(\bfx)&=&\bfsi^{\Gamma}{\bf n}^{\Gamma}(\bfx), \ \
 \bfsi^{\Gamma}={\rm const.},\ \bfx\in\Gamma,
\EEEQ
where $\bfep^{\Gamma}(\bfx)={1\over 2}\big[\nabla\otimes\bfu^{\Gamma}
(\bfx)+(\nabla\otimes
\bfu^{\Gamma}(\bfx))^{\top}\big],\ \bfx\in \Gamma$, and $\bfep^{\Gamma}$
and $\bfsi^{\Gamma}$ are the macroscopic strain and stress tensors, i.e. the given constant symmetric tensors.

It is assumed that the representative macrodomain $w$ contains a statistically
large number of realizations $\alpha$ of inclusions $v_i\in v$ (providing validity of the standard probability theory technique)
of the constituent $v_i\in v$
$(i=1,2,\ldots)$. A random parameter $\alpha$ belongs to a sample space ${\cal A}$, over which a probability density $p(\alpha)$ is defined
(see, e.g., [7]).
For any given $\alpha$, any random function $\bfg(\bfx,\alpha)$ (e.g., $\bfg=V,\bfsi,\bfep$) is defined explicitly as one particular member, with label $\alpha$, of an ensample realization. Then, the mean, or ensemble average  is defined by the angle brackets enclosing the quantity $\bfg$
\BB
\lle\bfg\rle(\bfx)=\int_{\cal A}\bfg(\bfx,\alpha)p(\alpha)d\alpha. 
\EE
No confusion will arise below in notation of the random quantity $\bfg(\bfx,\alpha)$ if the label $\alpha$ is dropped for compactness of expressions unless such indication is necessary.
One treats two material length scales (see, e.g., [8]):
the macroscopic scale $L$, characterizing the extent of $w$, and the microscopic scale $a$, related with the heterogeneities $v_i$.  Moreover, one supposes that applied field varies on a characteristic length scale $\Lambda$. The limit of our interests for both the material scales and field one  is presented in an asymptotic sense
\BB
L\gg\Lambda \ge a, 
\EE
as the scale of microstructure $a$ relative to the macroscale $L$
tends to zero.
All the random quantities  under
discussion are described by statistically inhomogeneous  random  fields.
    For the alternative description of the random structure
of a composite material let us introduce
 a conditional probability density $\varphi (v_i,{\bf x}_i
\vert v_1,{\bf x}_1, \ldots,v_n,{\bf x}_n)$,
which is a probability density to find the
$i$-th inclusion  with the center ${\bf x}_i$ in the domain
$v_i$ with fixed inclusions  $v_1,\ldots,v_n$
with the centers ${\bf x}_1,\ldots ,{\bf x}_n$. The notation
 $\varphi (v_i  , {\bf x}_i\vert ;v_1,{\bf x}_1,\ldots ,v_n,{\bf x}_n)$ denotes
the case ${\bf x}_i\neq {\bf x}_1,\ldots ,{\bf x}_n$. We will consider a general case of
statistically inhomogeneous media { with the homogeneous matrix} (for example for so-called Functionally Graded Materials (FGM)),
when the conditional probability density is  not invariant
with respect to translation:  $\varphi (v_i
 , {\bf x}_i\vert v_1,{\bf x}_1,\ldots ,v_n,{\bf x}_n)$ $\neq
 \varphi (v_i, {\bf x}_i+\bfx\vert v_1,{\bf x}_1+{\bf x},\ldots ,v_n,
{\bf x}_n+{\bf x})$, i.e. the microstructure functions depend upon
their absolute positions.
In particular, a random medium is called statistically homogeneous
in a narrow sense if its multi-point  statistical moments of
any order are shift-invariant functions of spatial variables.
Of course, $\varphi(v_i, {\bf x}_i\vert ;v_1,{\bf x}_1,\ldots
,v_n,{\bf x}_n)=0$ for values of ${\bf x}_i$ lying
inside the ``excluded volumes''
$\cup v^0_{mi}$ (since inclusions cannot overlap, $m=1, \ldots ,n)$, where  $v^0_{mi}\supset v_m$ with characteristic function $V^0_{mi}$
is the ``excluded volumes'' of $\bfx_i$ with respect to $v_m$
(it is usually assumed that $v^0_{mi}\equiv v^0_m$), and
 $\varphi
(v_i, {\bf x}_i\vert ;v_1,{\bf x}_1,\ldots ,v_n,{\bf x}_n)\to \varphi(v_i,
{\bf x}_i)$
as $\vert {\bf x}_i-{\bf x}_m\vert\to \infty ,
\ m=1,\ldots,n$ (since no long-range order is assumed).
$\varphi (v_i,{\bf x})$
is a number density, $n=n({\bf x})$ of component $v\ni v_i$
at the point ${\bf x}$  and $c^{(1)}({\bf x})$ is
the
concentration, i.e. volume fraction, of the component $v_i\in v$ at the point
 ${\bf x}$:
$ c^{(1)}({\bf x})=\langle V\rangle ({\bf x})=\overline v_in^{(1)}({\bf x}),
\ \overline v_i={\rm mes} v_i;\ i=1,2,\ldots),\quad
c^{(0)}({\bf x})=1-\langle V\rangle ({\bf x}).$
 Hereafter  only if the pair distribution
 function $g({\bf x}_i-{\bf x}_m)\equiv
 \varphi(v_i, {\bf x}_i\vert; v_m,{\bf x}_m)/n^{(k)}$
 depends on $\vert {\bf x}_m-{\bf x}_i\vert$  it is called the
 radial distribution function (RDF).
  The notations $\langle (.)\rangle ({\bf x})$ and
 $\langle (.)\vert v_1,{\bf x}_1;\ldots ;v_n,{\bf x}_n\rangle ({\bf x})$
 will be used for the average and for the conditional average taken
for the ensemble of a statistically inhomogeneous
 field $X=(v_i)$ at the point ${\bf x}$,
on the condition that there are inclusions at
 the points ${\bf x}_1,\ldots,{\bf x}_n$ and
${\bf x}_i\neq\bfx_j$ if $i\neq j$ ($i,j=1,\ldots, n)$.
  The notations $\langle (.)\vert; v_1,{\bf x}_1;\ldots;v_n,{\bf x}_n\rangle ({\bf x})$
 are used for   the case ${\bf x}\notin v_1,\ldots,v_n$.
The notation $\lle(\cdot)\rle_i(\bfx)$ at $\bfx\in v_i$ means the average over an ensemble realization of surrounding inclusions  (but not over the volume $v_i$ of a particular inhomogeneity, in contrast to $\lle(\cdot)\rle_{(i)}$) at the fixed $v_i$.
Without loss of generality, we assume that the subdomains $v^{(1)}$  do not touch the boundary $\Gamma$; such subdomains are called floating subdomains. In other words, the body $w$ is considered 
as one cut out from an infinite random medium and the inclusions $v_i$ intersected with the boundary $\Gamma$ are replaced by the matrix material. 

We will use two sorts of conditional averages of some tensor $\bfg$ (e.g., $\bfg=\bfep, \bfsi$).  
At first, the conditional statistical average in the inclusion 
phase $\lle\bfg \rle^{(q)}(\bfx)$ $\equiv\lle\bfg V\rle^{(q)}(\bfx)$ (at the condition that the point 
$\bfx$ is located in the inclusion phase $\bfx\in v^{(q)}$) can be found as
$\lle\bfg V\rle^{(q)}(\bfx)=\lle V^{(q)}(\bfx)\rle^{-1}\lle\bfg V^{(q)}\rle(\bfx)$. 
Usually, it is simpler to estimate the second conditional averages 
of these tensors in the concrete point $\bfx$ of the fixed inclusion $\bfx\in v_q$:
$\lle\bfg| v_q,\bfx_q\rle(\bfx)\equiv \lle \bfg\rle_q(\bfx)$.
At first we built some auxiliary set $v_q^1(\bfx)$
 with the boundary $\partial v^1_q({\bf x})$ formed by the centers of translated ellipsoids $v_q({\bf 0})$  
around the fixed point $\bfx$. 
 We construct $v_q^1(\bfx)$ as a limit $v_{kq}^0\to v_q^1({\bf x})$ if a fixed ellipsoid $v_k$ is shrinking to the point $\bfx$. Then we can get a relation between the mentioned averages [$\bfx=(x_1,\ldots,x_d)^{\top}$]:
\BB
\lle  \bfg\rle^{(q)}(\bfx)=\int_{ v_q^1({\bf x})}
n^{(q)}(\bfy)\lle \bfg|v_q,\bfy\rle(\bfx) ~d\bfy.  
\EE
 Formula (2.11) is valid for any material inhomogeneity of inclusions of any concentration in the macrodomain $w$ of any shape (if $v_q^1({\bf x})\subset w$). Obviously, the general Eq. (2.11) is reduced to the popular one 
$\lle\bfg\rle^{(q)}=\lle\bfg\rle_q$  for statistically homogeneous media subjected to homogeneous boundary conditions. 

\vspace{15pt}
\noindent{\bf 3 General integral equation}
\bigskip

\setcounter{equation}{0}
\renewcommand{\theequation}{3.\arabic{equation}}
Substituting the constitutive equation (2.1) and the Cauchy equation (2.3) into
the equilibrium equation (2.1), we obtain a differential
equation with respect to the displacement $\bfu$ which
can be reduced to  a symmetrized integral form after integrating by parts
(see, e.g, 
Chapter 7 in [1])
\BBEQ
{\bfep}(\bf x)&=&{\bfep}^0(\bfx)
 + \int_{w} {\bf U}(\bfx-\bfy)\bftau(\bfx)
d{\bf y},
\EEEQ
 where $\bftau(\bfx)\equiv{\bfL}_1({\bf y}){\bfep}({\bf y})$ is called the stress polarization tensor, and  the surface integral is absent because the heterogeneous are assumed (without loss of generality) to be floating ones. The integral operator kernel ${\bf U}$
is an even homogeneous a generalized function of degree $-d$ defined by the
second derivative of the Green tensor ${\bf G}$:
$ U_{ijkl}(\bfx)=\big[\nabla_j\nabla_l G_{ik}(\bfx)\big]_{(ij)(kl)}$,
the parentheses in indices mean symmetrization.
${\bf G}$
is the infinite-homogeneous-body Green's function of the
 Navier equation with an elastic modulus ${\bfL}^{(0)}$
defined by $
\nabla \left\lbrace{{\bfL}^{(0)}
\left[{\nabla \otimes {\bf G}^{(0)}(\bfx)+
 (\nabla\otimes{\bf G}^{(0)}(\bfx))^{\top}}\right]/2}\right\rbrace =-\bfdel\delta
 ({\bf x}),$
and vanishing at infinity ($|\bfx|\to\infty$),
$\delta ({\bf x})$ is the Dirac delta function.  The deterministic function
${\bfep}^0(\bfx)$ is the strain field which would  exist in the medium
with homogeneous properties ${\bfL}^{(0)}$ and appropriate
boundary conditions (see, e.g, [9]):
\BBEQ
{\varepsilon}_{pq}^0(\bfx)=\int_{\Gamma}
\Big[G_{i(p,q)}(\bfx\!-\!\bfs){t}_{i}^0(\bfs)-
u_i^0(\bfs)L^{(0)}_{ijkl}G^{(0)}_{k(p,q)l}(\bfx-\bfs)
n_j(\bfs)\Big]d\bfs 
,
\EEEQ
which conforms with the stress field $\bfsi^0(\bfx)={\bfL}^{(0)}\bfep^0(\bfx)$.
The representation (3.2) is valid for both the general cases of the first and
second boundary value problems as well as for the mixed boundary-value problem
(see for references [1]).
 For simplicity we will consider only internal points
$\bfx\in w$ of the microinhomogeneous macrodomain $w$ at sufficient distance from the boundary
\BB
a\ll|\bfx-\bfs|,\ \forall \bfs\in \Gamma. 
\EE
In so doing, some Cauchy data $[\bfu^0(\bfs),\bft^0(\bfs)]$  (3.2) (if they are not prescribed by the boundary conditions) will depend on perturbations introduced by all inhomogeneities, and, therefore $\bfep^0(\bfx)=\bfep^0(\bfx,\alpha)$.

Now we will center Eq. (3.1),
i.e. from both sides of Eq. (3.1) their statistical averages are subtracted
\BBEQ
\!\!\!\!\!\!\!\!\!\!{\bfep}({\bf x})
\!\!\!\!&=&\!\!\!\!
\langle {\bfep}\rangle ({\bf x})
\!+\!\!\int_{w}[\bfU(\bfx-\bfy)\bftau(\bfy)\!-\!\lle\bfU(\bfx-\bfy)\bftau\rle(\bfy)]
 d{\bf y}
+\bfcI^{\Gamma}_{\epsilon}
. 
\EEEQ
Without loss of generality, it is assumed
the traction boundary conditions (2.6). Then
$ 
\bfcI^{\Gamma}_{\epsilon} \equiv  \bfep^{0}(\bfx,\alpha)-\lle\bfep^0\rle(\bfx) 
$
defined by the integral (3.2) (containing only $\bfu^0(\bfs)$) over the external surface $\bfs\in\Gamma$
can be  dropped out, because  this tensor vanishes
 at sufficient distance ${\bf x}$
from the boundary $\Gamma$ (3.3) (see for details,
 e.g., Ref. [10]  
and its applications {Shermergor} [11]; see also [1], [12] where the case of nonfloating subdomains is considered).

The integrals in Eqs. (3.4) converges absolutely for both the statistically homogeneous and inhomogeneous random fields $X$ of inhomogeneities.
 Indeed, even for the FGMs, the term in the square brackets in Eq. (3.4) is of order $O(\vert {\bf x-y}\vert^{-2d})$  as $\vert {\bf x-y}
\vert \to\infty$, and the  integral in Eq. (3.4) 
converges absolutely. 
Therefore, for $\bfx\in w$ considered in Eqs. (3.4) and removed far enough from the boundary $\Gamma$ (3.3),
the right-hand side integrals in Eq. (3.4) does not depend
on the shape and  size of the domain $w$, and it can be replaced by the
integrals over the whole space $R^d$. With this assumption we hereafter
omit explicitly denoting $R^d$ as the integration domain in the equation
\BBEQ
\!\!\!\!\!\!{\bfep}({\bf x})
\!\!\!\!&=&\!\!\!\!
\langle {\bfep}\rangle ({\bf x})
\!+\!\int [{\bf U}(\bfx-\bfy)\bftau(\bfy)\!-\!
\lle\bfU(\bfx-\bfy)\tau(\bfy)\rle(\bfy)]d\bfy.
\EEEQ

It should be mentioned that a popular equality 
\BB
\lle\bfU(\bfx-\bfy)\tau(\bfy)\rle(\bfy)=\bfU(\bfx-\bfy)\lle\tau(\bfy)\rle(\bfy) 
\EE
is only asymptotically valid at $|\bfx-\bfy|\to \infty$. Then Eq. (3.5) is asymptotically reduced to the known one (see for details [1])
\BBEQ
{\bfep}({\bf x})
\!\!\!\!&=&\!\!\!\!
\langle {\bfep}\rangle ({\bf x})
+\int {\bf U}(\bfx-\bfy)[\bftau(\bfy)-
\lle\tau\rle(\bfy)]d\bfy,
\EEEQ
which in turn coincides with the equation
\BBEQ
{\bfep}({\bf x})
\!\!\!\!&=&\!\!\!\!
\langle {\bfep}\rangle 
+\int {\bf U}(\bfx-\bfy)[\bftau(\bfy)-
\lle\tau\rle]d\bfy,
\EEEQ
for statistically homogeneous media subjected to the homogeneous boundary conditions.

Let the inclusions  $v_1,\ldots,v_n$ be fixed
and we define two sorts of effective fields
$\overline{\bfep}_i(\bfx)$ and  $\widetilde{\bfep}
_{1,\ldots,n}(\bfx)\quad (i=1,\ldots,n;\ {\bf x}\in v_1,\ldots,v_n)$
by the use of the rearrangement of Eq. (3.6) in the following
form (see for the earliest references of related manipulations [1]):
\BBEQ
\bfep(\bfx)\!\!\!&=&\!\!\!\overline{\bfep}_i(\bfx)+\int{\bf U}(\bfx-\bfy)V_i(\bfy)
\bftau(\bfy)d\bfy,\nonumber\\ 
\overline{\bfep}_i(\bfx)\!\!\!&=&\!\!\!\widetilde {\bfep}
_{1,\ldots,n}(\bfx)+\sum_{j\neq i}\int
{\bf U (x-y)}V_j({\bf y})\bftau({\bf y})d{\bf y},\nonumber\\ 
\widetilde {\bfep}_{1,\ldots,n}(\bfx)
\!\!\!\!&=&\!\!\!\!\langle \bfep\rangle ({\bf x})
+\int \Big\{{\bf U (x-y)}  \bftau(\bfy)
V(\bfy\vert;v_1,{\bf x}_1;\ldots;v_n,{\bf x}_n)
\nonumber\\
\!\!\!\!&-&\!\!\!\!
\langle \bfU(\bfx-\bfy)\bftau\rangle ({\bf y})\Big\} d{\bf y},
\EEEQ
 for  ${\bf x}\in v_i,\ i=1,2,\ldots,n$; here 
$V(\bfy\vert v_1,{\bf x}_1;\ldots ;v_n,{\bf x}_n)$ is
   a random characteristic function of inclusions $\bfx\in v$ under
   the condition that ${\bf x}_i\neq\bfx_j$ if $i\neq j$ ($i,j=1,\ldots, n)$.
Then, considering  some conditional statistical averages of
the general integral equation (3.5) leads
      to   an infinite system of new integral equations $(n=1,2,\ldots)$
\BBEQ
\langle \bfep \vert v_1,{\bf x}_1;\!\!\!\!&\ldots&\!\!\!\! ;v_n,{\bf x}_n
\rangle ({\bf x})
-\sum^n_{i=1}\int{\bf U (x-y)}\langle V_i({\bf y})
\bftau \vert v_1,{\bf x}_1;\ldots;v_n,{\bf x}_n\rangle ({\bf y})
d{\bf y}\nonumber\\
\!\!\!\!&=&\!\!\!\! \langle \bfep\rangle ({\bf x})
+\int \Big\{{\bf U (x-y)}\langle \bftau
\vert;v_1,{\bf x}_1;\ldots;v_n,{\bf x}_n\rangle ({\bf y})
\nonumber\\\!\!\!\!&-&\!\!\!\!
\lle \bfU(\bfx-\bfy)\bftau\rangle ({\bf y})\Big\}d{\bf y}.
\EEEQ
Since ${\bf x}\in v_1,\ldots,v_n$
in the $n$-th line of the system can take the values of the 
inclusions $v_1,\ldots,v_n$, the $n$-th line actually contains $n$ equations.

\vspace{15pt}
\noindent{\bf  
 4 Background of analytical micromechanics}
\bigskip

\noindent{\it  4.1 Approximate effective field hypothesis}
\bigskip

\setcounter{equation}{0}
\renewcommand{\theequation}{4.\arabic{equation}}
In the current section 4, only statistically homogeneous media subjected to homogeneous boundary conditions (2.7) are considered.
 In order  to simplify the exact system (3.10) we now apply the 
so-called effective field hypothesis which is the main approximate hypothesis of many micromechanical methods:

\vspace{10pt}
{\bf Hypothesis 1a, H1a}. {\sl Each inclusion $v_i$  is located
in the field (3.9$_2$)}
\BB
\overline {\bfep}_i({\bf y})
\equiv \overline {\bfep}({\bf x}_i)\ (\bfy\in v_i),
\EE
{\sl which is homogeneous
over the inclusion $v_i$}.
\vspace{10pt}

In some methods (such as, e.g., the MEFM) this basic hypothesis {\bf H1a} is complimented by a satellite hypothesis [compare with (3.6)]:

\vspace{10pt}
{\bf Hypothesis 1b, H1b.}
{\sl The perturbation introduced by the inclusion $v_i$
at the point ${\bf y}\notin v_i$ is defined by the relation}
\BB
\int{\bf U(y-x)}V_i({\bf x})\bftau({\bf x})d{\bf x}=
\bar v_i{\bf T}_i{\bf (y-x}_i)
\bftau_i.
\EE
\vspace{10pt}

Hereafter $\bftau_i\equiv \langle \bftau({\bf x})V_i({\bf x})
\rangle _{(i)}$ is an average over
the  volume of the inclusion $v_i$ (but not over the ensemble),
$\langle(.)\rangle_i\equiv \langle\langle(.)
\rangle_{(i)}\rangle$, and ($\bfx_i,\bfz\in v_i,\ \bfx\not\in v_i,\ \bfx_j,\bfy\in v_j$)
\BBEQ
\bfT_i{\bf ( x \!\! - \!\! x}_i)=
\lle\bfU(\bfx-\bfz)\rle_{(i)}, \ \ \
{\bf T}_{ij}(\bfx_j-\bfx_i)=\langle{\bfT}_i(\bfy-\bfx_i)\rangle_{(j)}. 
\EEEQ
If $\bfx\in v_i$ then $\bfT_i{\bf ( x  -  x}_i)=-(\overline v_i)^{-1}\bfP_i\equiv$const., 
where the  tensor ${\bf P}_i$ is associated with the 
well-known Eshelby tensor by $\bfS_i={\bfP_i}\bfL^{(0)}$. 
For a homogeneous ellipsoidal inclusion $v_i$ the standard assumption (4.1) 
(see, e.g., [1]) 
yields  the assumption (4.2), otherwise the formula 
(4.2) defines an additional assumption. 
The tensors ${\bf T}_{ij}({\bf x}_i-{\bf x}_j)$ has an analytical 
representation for spherical inclusions of different size in an  isotropic
matrix (see for references [1]).

According to hypothesis ${\bf H1a}$ and
 in view of the linearity of the problem  there exist 
constant fourth and second-rank tensors ${\bf A}_i({\bf x}),\ 
{\bf R}_i{\bf (x)}$, such that
\BB\bfep ({\bf x})={\bf A}_i{\bf (x)}\overline 
{\bfep} ({\bf x}_i),\quad
\bftau({\bf x})
={\bf R}{\bf (x)}\overline {\bfep} ({\bf x}_i),
\quad {\bf x}\in v_i,
\EE
where $v_i\subset v^{(i)}$ and
${\bf R}{\bf (x)}={\bf L}_1^{(1)}{\bf A}_i{\bf 
(x)}$.
According to {Eshelby}'s [13]
theorem there are the following relations between the 
averaged tensors (4.4)
${\bf R}=\overline v_i
{\bf P}_i ^{-1}({\bf I-A}_i),$
where ${\bf g}_i\equiv \langle {\bf g(x)}\rangle _{(i)} \quad ({\bf g}$ stands for $\bfA_i,\bfR)$. For
example, for the homogeneous ellipsoidal domain $v_i$ (2.4) 
 we obtain ${\bf A}_i=\left({{\bf I+P}_i{\bf L}_1^{(i)}}\right)^{-1}$. 
In the general case of coated inclusions $v_i$, the tensors 
$\bfA_i(\bfx)$ can be found by the transformation method by 
 Dvorak  and {Benveniste} [14] 
(see for references and details [1]).

\bigskip
\noindent{\it
4.2 Closing hypothesis}
\bigskip

For termination of the  hierarchy of statistical moment equations (3.10)
we will use the closing EFH
called the 
``quasi-crystalline" approximation by Lax [15] 
which in our notations has a form 

\vspace{10pt}
{\bf Hypothesis 2, ``quasi-crystalline" approximation}.
{\sl It is supposed that the mean value of the effective field at a point
$\bfx\in v_i$ does not depend on the stress field  inside surrounding heterogeneities  
$v_j\not = v_i$}: 
\BB
\langle \overline{\bfep}_i({\bf x})\vert v_i,{\bf x}_i;v_j,{\bf x}_j \rangle = \langle \overline
{\bfep}_i \rangle ,\quad {\bf x}\in v_i. 
\EE
\vspace{10pt}

In the framework of the hypothesis {\bf H1}, substitution  of the solution
(4.4) into the first equation of 
 the system (3.9) at $n=1$ and  at the EFH
{\bf H2} 
leads to the solution ($\bfx\in v_i$)
\BBEQ
\langle  \overline{\bfep}  \rangle _i
\!\!\!\!&=&\!\!\!\!
\bfR^{-1}{\bf Y}{\bf R} \langle \bfep \rangle ,\\ 
\langle  {\bftau}  \rangle _i(\bfx)
\!\!\!\!&=&\!\!\!\!
\bfR(\bfx)\bfR^{-1}{\bf Y}{\bf R} \langle \bfep \rangle,\\ 
\bfL^*\!\!\!\!&=&\!\!\!\!\bfL^{(0)}+\bfY\bfR c^{(1)},  
\EEEQ
 where the matrix ${\bf Y}$ determines the action of the 
surrounding inclusions on the considered one and has an inverse  matrix 
${\bf Y}^{-1}$ given by
\BBEQ
({\bf Y}^{-1})=
\bfI-\bar v_i\bfR
\!\!\int \!\Big[{\bf T}_{iq}({\bf x}_i-{\bf x}_q)
  \varphi (v_q,{\bf x}_q\vert ;v_i,{\bf x}_i) 
-{\bf T}_i({\bf x}_i-
  {\bf x}_q)n^{(1)}\Big]d{\bf x}_q.
\EEEQ
General case of the closing hypothesis taking $n$ interacting heterogeneities is considered in Chapter 8 in [1].

\bigskip
\noindent{\it 4.3 Hypothesis of ``{\it ellipsoidal symmetry}" of composite structure}
\bigskip

To make further progress, the hypothesis of ``{\it ellipsoidal symmetry}" for the distribution of inclusions attributed to Willis [5]
is widely used:

\vspace{10pt} 
{\bf Hypothesis 3, ``ellipsoidal symmetry"}.  
{\sl The conditional probability density function $\varphi (v_{j},{\bf x}_j \mid ;v_{i},{\bf x}_{i})$ depends on $\bfx_j-\bfx_i$ only through the combination $\rho=|({\bf a}^0_{ij})^{-1} ({\bf x}_{j}-{\bf x}_{i})|$}:
\BB
\varphi (v_{j},{\bf x}_j \mid ;v_{i},{\bf x}_{i})
=h
(\rho ),\ \ \  \rho \equiv \mid 
({\bf a}^0_{ij})^{-1} ({\bf x}_{j}-{\bf x}_{i})\mid
\EE
{\sl where the matrix $({\bf a}^0_{ij})^{-1}$ (which is symmetric in the indexes $i$ and 
$j$, ${\bf a}^0_{ij}={\bf a}^0_{ji}$)
defines the ellipsoid excluded volume $v^0_{ij}=\{\bfx:\ |({\bf a}^0_{ij})^{-1}\bfx|^2< 1\}$.}
\vspace{10pt} 

For  spherical inclusions the 
relation  (4.10) is realized for a statistical isotropy of the composite 
structure. It is reasonable to assume that 
$({\bf a}^0_{ij}) ^{-1}$ identifies
a matrix of affine transformation that transfers  the ellipsoid
$v_{ij}^0$ being the ``excluded volume" (``correlation hole") into a  unit sphere and, therefore, the representation of the matrix $\bfY$ can be simplified: 
\BB
({\bf Y}^{-1})={\bf I}-c^{(i)}{\bf R}\bfP_{i}^0,
\EE
where for the sake of simplicity of the subsequent calculation we will usually assume that the shape of ``correlation hole" $v^0_{ij}$ does not depend 
on the inclusion $v_j$: $v^0_{ij}=v_i^0$ and $\bfP_{ij}^0=\bfP_i^0\equiv \bfP(v_i^0)$. 
 
The concept of the EFH (even if this term is not mentioned) in combination with subsequent assumptions (e.g., mentioned above) totally dominates (and creates the fundamental limitations) in all four groups of analytical micromechanics in physics and mechanics of heterogeneous media: model methods, perturbation methods, self-consistent methods (e.g., Mori-Tanaka, MT, approach, and the MEFM), and variational ones (see for references [1]).

\vspace{10pt}
\noindent{\bf  
5 Background of computational analytical micromechanics}
\bigskip

\noindent {\it 5.1 A single inclusion subjected to inhomogeneous prescribed effective field}
\medskip

\setcounter{equation}{0}
\renewcommand{\theequation}{5.\arabic{equation}}
In the current subsection we will consider a satellite problem whose solution will be used for estimation of effective properties of composites in Subsection 5.2. Namely, let the inclusions  $v_i$ be fixed
and loaded by the inhomogeneous effective field 
$\overline{\bfep}_i(\bfx)$. Then we used the known regularized integral equation
\BB
\bftau ({\bf x})=\overline{\bftau}_i(\bfx)+
\int \bfK_i (\bfx,\bfy)
\big[\bftau({\bf y})-\bftau(\bfx)\big]  
 d{\bf y,} \quad \bfx\in v_i,
\EE
where $\overline{\bftau}_i(\bfx)={\bf E}_i(\bfx)\overline{\bfep}(\bfx)$, $(\bfx\in v_i$) is called
the effective stress polarization tensor in the inclusion $v_i$,
and (no sum on $i$)
\BBEQ
{\bf K}_i(\bfx,\bfy)&=&{\bf E}_i(\bfx)
{\bf U}(\bfx-\bfy)V_i(\bfy), \\
{\bf E}_i(\bfx)&=&{\bf L}_1(\bfx)
[{\bf I}+{\bf P}^0_i(\bfx){\bf L}_1(\bfx)]^{-1}.
\EEEQ
Here the tensor $\bfP_i^0(\bfx)=-\int V^0_i(\bfy){\bf U (x-y)}d{\bf y}$ can be estimated, e.g., by the FEA and assumed to be known.

We formally write the solution of Eq. (5.1)
as
\BB
\bftau=\bfcL_i\!*\!\overline{\bftau}_i,
\EE 
where the inverse operator $\bfcL_i=({\bf I}-\bfcK_i)^{-1}$ will be 
constructed by
the iteration method based on the  recursion formula
\BB
\bftau^{[k+1]}=\overline{\bftau}_i+\bfcK_i\bftau^{[k]},
\EE
the convergence of which is analyzed in [1].
Here the integral operator $\bfcK_i$ has the kernel formally represented as
\BB
\bfcK_i(\bfx,\bfy)={\bf K}_i(\bfx,\bfy)-\delta(\bfx-\bfy)\int
V_i(\bfz){\bf K}_i(\bfx,\bfz)d\bfz,
\EE
and one used
an initial approximation
\BB
{\bftau}^{[0]}
(\bfx)=\overline{\bftau}_i(\bfx), 
\EE
which is exact for a homogeneous ellipsoidal inclusion subjected to 
 remote homogeneous stress field $\overline{\bfep}(\bfx)\equiv \overline{\bfep}={\rm const.}$

The solution (5.4) allows us to state that 
the linear operators $\bfcL^{\epsilon}$ and $\bfcL^{\tau}$ describing a perturbation of the strain fields inside and outside the inclusion $v_i$ ($\bfx\in R^d$)
\BBEQ
\!\!\!\!\!\!\!\!\!\!\!\!\int\!\!{\bf U}(\bfx\!-\!\bfy)V_i(\bfy)
\bftau(\bfy)d\bfy\!\!\!\!&=&\!\!\!\!\bfep(\bfx)\!-\!\overline{\bfep}_i(\bfx)\!\!\equiv\!\!
\bfcL_i^{\epsilon}(\overline{\bfep}_i)(\bfx)\!\!\equiv\!\! \bfcL^{\tau}_i(\bftau)(\bfx),\\ 
\bfcL_i^{\epsilon}(\overline{\epsilon}_i)(\bfx)
\!\!\!\!&=&\!\!\!\! \int{\bf U}(\bfx-\bfy)
\bfcL_i*(\bfE_i\overline{\bfep})(\bfy)V_i(\bfy) d\bfy, \\
\bfcL_i^{\tau}(\bftau)(\bfx)
\!\!\!\!&=&\!\!\!\! \int{\bf U}(\bfx-\bfy)\bftau(\bfy)V_i(\bfy)
d\bfy.  
\EEEQ
are constructed.

\bigskip
\noindent{\it 5.2 Estimation of effective elastic moduli} 
\medskip

The new general integral equation (3.5) 
can be rewritten in terms of the operator representation $\bfcL^{\tau}$ (5.8)
\BB
\bfep(\bfx)=\lle\bfep\rle(\bfx)+\int[\bfcL^{\tau}(\bftau)(\bfx)-
\lle\bfcL^{\tau}(\bftau)\rle(\bfx)]d\bfy. 
\EE

For statistically homogeneous media subjected to homogeneous boundary conditions (2.7) and in the framework of the quasi-crystalline approximation (4.5),
 conditional averaging of Eq. (5.11) can be solved by the iteration method 
\BBEQ
\lle\overline{\bfep}\rle_{i}^{[n+1]}(\bfx)\!&=&\!\lle\bfep\rle+\int
\bfcL^{\tau}_q(\lle{\bftau}\rle_{q}^{[n]})(\bfx)
[\varphi(v_q,\bfx_q|;v_i,\bfx_i)-n^{(q)}(\bfx_q)]d\bfx_q,\nonumber\\
\lle{\bftau}\rle_{q}^{[n+1]}(\bfx)&=&\bfcR_q*
\lle\overline{\bfep}\rle_{q}^{[n+1]}(\bfx),
\EEEQ
where $\bfcR_q=\bfL_1^{(1)}(\bfI+\bfcL^{\epsilon}_q)$.
Generalization of Eq. (5.12) to the cases of both the statistically inhomogeneous media and other multiparticle closing assumptions (see [1]) is obvious. 
A convergence of the sequence $\lle\bftau^{[n]}\rle_i(\bfx)$ (5.12) is analyzed analogously to the sequence (5.5).
An initial approximation $\lle{\bftau}\rle_{i}^{[0]}(\bfx)$ is defined by the classical approach (4.7) and (4.11). 
It suggests the Neumann series form for the  solution $\lle\overline{\bfep}\rle_{i}^{[n]}(\bfx)\to \lle\overline{\bfep}\rle_{i}(\bfx)$ 
(as $n\to\infty$) of (5.12) and $\lle{\bftau}\rle_{i}^{[n]}(\bfx)=
\bfcL_i*(\bfE_i \lle\overline{\bfep}\rle_{i}^{[n]})(\bfx)$:
\BB
\lle{\bftau}\rle_i(\bfx)\equiv\lim_{n\to \infty}\lle\bftau^{[n]}\rle_i(\bfx)=\bfR_i^*(\bfx)
\lle\bfep\rle, 
\EE 
which yields the final representations for the effective properties
\BB
\bfL^*=\bfL^{(0)}+\lle\bfR^*V\rle. 
\EE 

\vspace{10pt}
\noindent{\bf  
6 Qualitative comparison of the classical and new approaches}
\bigskip

The hypothesis {\bf H1} is widely used (explicitly or implicitly) for the majority of the methods of micromechanics even if the term ``effective field hypothesis" is not indicated. For example, Buryachenko [1] 
demonstrated that hypothesis {\bf H1} is exploited in the effective medium method, generalized self-consistent method, differential methods, Mori-Tanaka method, the MEFM, conditional moments method, variational methods, and others. These are a lot of other methods using the hypothesis {\bf H1} differ one from another by some additional specific assumptions used at the analysis  of the initial integral equations either Eqs. (3.5), (3.7), or (3.8).

The differences of Eqs. (3.5), (3.7), and (3.8)
are fundamental for subsequently solving the truncated hierarchy (3.10)  
 involving a rearrangement of each appropriate equation before it is solved. The most successful rearrangement are those which make the right-hand side of the coupled equations reflect the detailed corrections to that  basic physics. 
So,  Eq. (3.8) was obtained by subtracting the difficult state at infinity from equation (3.1), i.e. roughly speaking the constant force-dipole density  expressed through an alternative technique of the Green's function. This dictates the fundamental limitation of a possible generalization of Eq. (3.8) to both the FGMs and inhomogeneous boundary conditions. The mentioned deficiency of Eq. (3.8) was resolved by Eq. (3.7) which the renormalizing term provides an absolute convergence of the integral in Eq. (3.7) at $|\bfx-\bfy|\to \infty$ for the general cases of the FGMs. However, the same term in Eq. (3.7) is used in a short-range domain $|\bfx-\bfy|<3a$ in the vicinity of the point $\bfx\in w$. A fundamental deficiency of Eq. (3.7) is a dependence of the renormalizing term
$\bfU(\bfx-\bfy)\lle\bftau\rle(\bfy)$ [obtained in the framework of the 
asymptotic approximation (3.6)] only on the statistical average $\lle\bftau\rle(\bfy)$ while the renormalizing term
$\lle\bfU(\bfx-\bfy)\bftau\rle(\bfy)$ in Eq. (3.5) explicitly depends on  
on details distribution $\lle\bftau|v_q,\bfx_q\rle(\bfy)$ 
($\bfy\in v_q$).
What seems to be only a formal trick [abandoning the use of the approximations (3.6)] is in reality a new background of micromechanics
[defining a new field of micromechanics called computational analytical micromechanics, CAM]
which yields the discovery of fundamentally new effects even in the theory of statistically homogeneous media subjected to homogeneous boundary conditions. So, the final classical representation of the effective properties (4.8) depends only on the average strain concentrator factor
 $\bfA_i$ while the effective properties (5.14) implicitly depend on the inhomogeneous tensor  $\bfA_i(\bfx)$. 
Moreover, the detected dependence of the effective properties (5.14) on the detailed strain concentrator factors $\bfA_i(\bfx)$  rather than on the average values  $\bfA_i$ allows us to abandon the hypothesis {\bf H1b} whose accuracy is questionable for the noncanonical inclusions. In such a case the statistical average effective field  estimated by Eq. (5.12) is found to be inhomogeneous that discards the hypothesis {\bf H1a}. 
Thus, the CAM does not involve the hypotheses {\bf H1a}, {\bf H1b}, and {\bf H3} as contrasted to the classical analytical micromechanics. Only the closing assumption {\bf H2} (or its multiparticle generalizations, see [1]) is exploited
in CAM. 

It should be mentioned, that 
the domain of the operator $\bfcL^{\epsilon}_q(\lle\overline{\bfep}\rle_{q}^{[n]})(\bfx)$ (5.12) is a whole space $\bfx\in R^d$, and, because of this, 
some points of the area $\bfx\in v_i$ in Eq. (5.12) can be uncovered by the heterogeneities $v_q$ and, therefore, the effective strain 
$\lle\overline{\bfep}\rle_i^{[n+1]}(\bfx)$ (5.12) will depend on the strain perturbations 
$\bfcL^{\epsilon}_q(\lle\overline{\bfep}\rle_q^{[n]})(\bfx)$ in the vicinity $\bfx\in v_q^{\oplus}$ of the area $v_q$ rather than only on stress distributions in the inhomogeneity $v_q$. 
In particular, for well-stirred approximation of the binary correlation function $\varphi(v_q,\bfx_q|v_i,\bfx_i)$ of ellipsoidal inclusions, $v_q^{\oplus}$ is expressed by the Minkowski addition $v_q^{\oplus}=v_q^0\oplus v_i$ while for the spherical inclusions $v_q^{\oplus}=\{\bfx|\ |\bfx-\bfx_q|< 3a\}$. 
 Thus, we obtain a fundamental conclusion that effective moduli (5.14) in general  depend not only on the strain distribution inside the inhomogeneities but also on the strains in the vicinities of heterogeneities. Then the size of the excluded volume as well as the RDF  will impact on the effective field  (5.12) even in the framework of hypothesis {\bf H3}.  Indeed, if the radius of the excluded volume $v_i^0$ for the spherical inclusions increases  from $2a$ to $3a$ then the long distance of the influence zone $v_q^{\oplus}$ of the inhomogeneity $v_q$ on the effective field $\lle\overline{\bfep}\rle_{i}(\bfx)$ will increase from the value $|\bfx-\bfx_q|=3a$  to $|\bfx-\bfx_q|=4a$. A larger difference between 
the backgrounds (3.5) and (3.8)
was obtained   for composites reinforced by either nonellipsoidal or inhomogeneous inclusions demonstrating essentially inhomogeneous stress distribution inside isolated heterogeneities even in the framework of the hypothesis {\bf H1}. 
 
Quantitative estimations of the analyses presented above are under  progress for some particular cases of fiber composites and will be considered in other publications.

\bigskip \noindent
{\bf Acknowledgments:}
\medskip

This work was partially supported by the Visiting Professor Program of the University of Cagliari funded by Regione Autonoma della Sardegna and by the  Eppley Foundation for Research.
 
\bigskip
\noindent{\bf References}
\medskip
{\baselineskip=9pt
\parskip=1pt
\lrm

\hangindent=0.4cm\hangafter=1\noindent
[1] Buryachenko, V. A. (2007) {\tenit Micromechanics of Heterogeneous Materials}. Springer, NY.

\hangindent=0.4cm\hangafter=1\noindent
[2]
 Kr\"oner, E.  (1958)
Berechnung der elastischen Konstanten des Vielkristalls
aus den Konstanstanten des Einkristalls.
{\tenit Z. Physik.}, {\tenbf 151}, 504--518.

\hangindent=0.4cm\hangafter=1\noindent
[3] 
 Mori, T.,  Tanaka, K. (1973) Average stress in matrix and average elastic energy of materials with misfitting inclusions. {\tenit  Acta Metall}., {\tenbf 21}, 571--574

\hangindent=0.4cm\hangafter=1\noindent
[4] 
Mossotti, O.F.  (1850) Discussione analitica sul'influenza che l'azione di un mezzo dielettrico ha sulla distribuzione dell'electricit\'a alla superficie di pi\'u corpi elettrici disseminati in eso.    {\tenit Mem Mat Fis della Soc Ital di Sci in Modena,} {\tenbf      24}, 49--74.  

\hangindent=0.4cm\hangafter=1\noindent
[5] 
Willis,  J.R.  (1977)
Bounds and self-consistent estimates for the overall properties
of anisotropic composites.  {\tenit J. Mech. Phys. Solids}, {\tenbf 25}, 185--203. 

\hangindent=0.4cm\hangafter=1\noindent
[6] 
 Lekhnitskii, A.G.  (1963) {\tenit Theory of Elasticity of an Anisotropic Elastic Body}. Holder Day, San Francisco.

\hangindent=0.4cm\hangafter=1\noindent
[7]
{Willis,}  J.R. (1981)
Variational and related methods for the overall properties
of composites. {\tenit Advances in Applied Mechanics},
 {\tenbf 21}, 1--78.

\hangindent=0.4cm\hangafter=1\noindent
[8] 
Torquato, S.  (2002) {\tenit Random Heterogeneous Materials: Microstucture and Macroscopic
Properties}. Springer-Verlag, New York, Berlin.

\hangindent=0.4cm\hangafter=1\noindent
[9] 
 Brebbia, C.A.,  Telles,  J.C.F.,  Wrobel, L.C. (1984)
{\tenit Boundary Element Techniques}. Springer-Verlag, Berlin.

\hangindent=0.4cm\hangafter=1\noindent
[10] 
 {Filatov}, A.N., {Sharov}, L.V.  (1979)
{\tenit Integral Inequalities and the Theory of Nonlinear Oscillations.}
 Nauka, Moscow (In Russian).

\hangindent=0.4cm\hangafter=1\noindent
[11] 
{Shermergor,}  T.D.   (1977) { \tenit The Theory of Elasticity  of Microinhomogeneous Media}. Nauka, Moscow (In Russian).

\hangindent=0.4cm\hangafter=1\noindent
[12] 
Buryachenko, V.A. (2009) On some background of multiscale analysis of heterogeneous materials. {\tenit Proceeding of the 10th U.S. National Congress for Computational Mechanics}. Columbus, USA.

\hangindent=0.4cm\hangafter=1\noindent
[15] 
 Eshelby, J.D. (1957) The determination of the elastic field of an ellipsoidal inclusion, and related problems. {\tenit Proc. Roy. Soc. Lond.}, {\tenbf A241}, 376-–396.

\hangindent=0.4cm\hangafter=1\noindent
[16] 
Dvorak, G.J.,  Benveniste, Y.     (1992)
On transformation strains and uniform fields in multiphase elastic media.    
{\tenit    Proc.   Roy. Soc.  Lond.,} {\tenbf    A437}, 291--310.

\hangindent=0.4cm\hangafter=1\noindent
[17]Lax,   M.   (1952)   Multiple scattering of waves II. The effective
fields dense systems. {\tenit Phys. Rev.} {\tenbf  85}, 621--629.

}

\end{document}